\documentclass{statsoc}

\usepackage[a4paper]{geometry} % Added for rendering

\usepackage{natbib}
\usepackage{graphicx} %aggiunto
\usepackage{url} %%AGGIUNTO
\usepackage{amsmath}%%AGGIUNTO
\usepackage{amssymb}%%AGGIUNTO
\usepackage[table]{xcolor}%%AGGIUNTO
\usepackage{comment}%%AGGIUNTO 
\definecolor{light-gray}{gray}{.9}%%AGGIUNTO
\usepackage{etoolbox}
\makeatletter
\patchcmd{\@makecaption}
{\parbox}
{\advance\@tempdima-\fontdimen2} % decrease the width!
{}{}
\makeatother

\graphicspath{{img/}}

\title[]{
A Bayesian approach to { model local and temporal} heterogeneity in repeated cross-sectional health surveys}
\author[Stival \textit{et al.}]{Mattia Stival}
\address{Department of Economics, Ca' Foscari University of Venice,
    Venice, 30121, 
    Italy.}
\email{mattia.stival@unive.it}
\author[Stival \textit{et al.}]{Lorenzo Schiavon}
\address{Department of Economics, Ca' Foscari University of Venice,
    Venice, 30121, 
    Italy.}
\email{lorenzo.schiavon@unive.it}
\author[Stival \textit{et al.}]{Stefano Campostrini}
\address{Department of Economics, Ca' Foscari University of Venice,
    Venice, 30121, 
    Italy.}
\email{stefano.campostrini@unive.it}

\begin{document}

\begin{abstract}
In several countries, including Italy, a prominent approach to population health surveillance involves conducting repeated cross-sectional surveys at short intervals of time. These surveys gather information on the health status of individual respondents, including details on their behaviours, risk factors, and relevant socio-demographic information.
While the collected data undoubtedly provides valuable information, modelling such data presents several challenges. For instance, in health risk models, it is essential to consider behavioural information, { local and temporal dynamics}, and disease co-occurrence.
In response to these challenges, our work proposes a multivariate { temporal}
logistic model for chronic disease diagnoses {at local level}. Linear predictors are modelled using individual risk factor covariates and a latent individual propensity to diseases.
Leveraging a state space formulation of the model, we construct a framework in which temporal heterogeneity in regression coefficients is informed by exogenous information {at local level}, corresponding to different contextual risk factors that may affect %health and 
the occurrence of chronic diseases in different ways.
To explore the utility and the effectiveness of our method, we analyse behavioural and risk factor surveillance data collected in Italy (PASSI), 
which is well-known as a country characterised by high peculiar administrative, social and territorial diversities reflected on high variability in morbidity among population subgroups.
\end{abstract}

\section{Introduction}
\label{sec:intro}
\noindent In the context of health risk factor studies, cross-sectional survey data represents a remarkable source of information and insights, particularly when these are repeated at short intervals over time \citep{campostrini2005institutionalization}. 
These surveys offer a practical alternative to longitudinal studies, 
reducing the costs of interviewing the same individuals across several years and avoiding the issue of censoring and sample size decreasing due to drop-out or death of the participants.
On the other hand, cross-sectional surveys provide snapshots of a population at specific moments in time only, thereby disallowing the study of population dynamics, which is a crucial aspect in health monitoring.
To face this trade-off, several countries, including Italy, have set health monitoring systems conducting a continuous data collection through a series (typically based on monthly samples) of cross-sectional surveys on the model of US Behavioural Risk Factor Surveillance System \citep[BRFSS,][]{nelson2001reliability}.
BRFSS data are collected by interviewing every year about $400,000$ subjects in the US population selected according to a random sample scheme that guarantees local representation. 
Questions regard health status, administrative information, behaviours and possible risk factors.
The data collected are mainly used to produce reports and indicators that describe, at various levels of aggregation, the health and behavioural risk status of the states.
While considering the distribution of answers to each question in itself may be of interest to policymakers, understanding the relationships between risk factors and health-related outcomes would offer a better understanding of how targeted policies can affect the well-being of population subgroups. 

Despite the wealth of knowledge collected by these systems, the existing literature often overlooks the potential depth and significance of the data, as, for instance, focusing only on an univariate risk or ignoring the spatio-temporal dynamics and the information about habits and behaviours.
Therefore, in order to fully tap into the data potential, we introduce a novel methodology to assess the risk of suffering one or multiple chronic diseases based on individual-specific risk factors.
Despite the widespread presence of models with similar aims in the existing BRFSS literature \citep{AssafJRSSA15, Pastore-passi22, ZHENG202374}, our approach introduces several original elements in the statistical specification, aiming to further unleash the information included in the data to explain the dynamics of the local populations.

\begin{figure}[ht]
 \label{fig:bias}
 \centering
             \includegraphics[width = 0.85\textwidth]{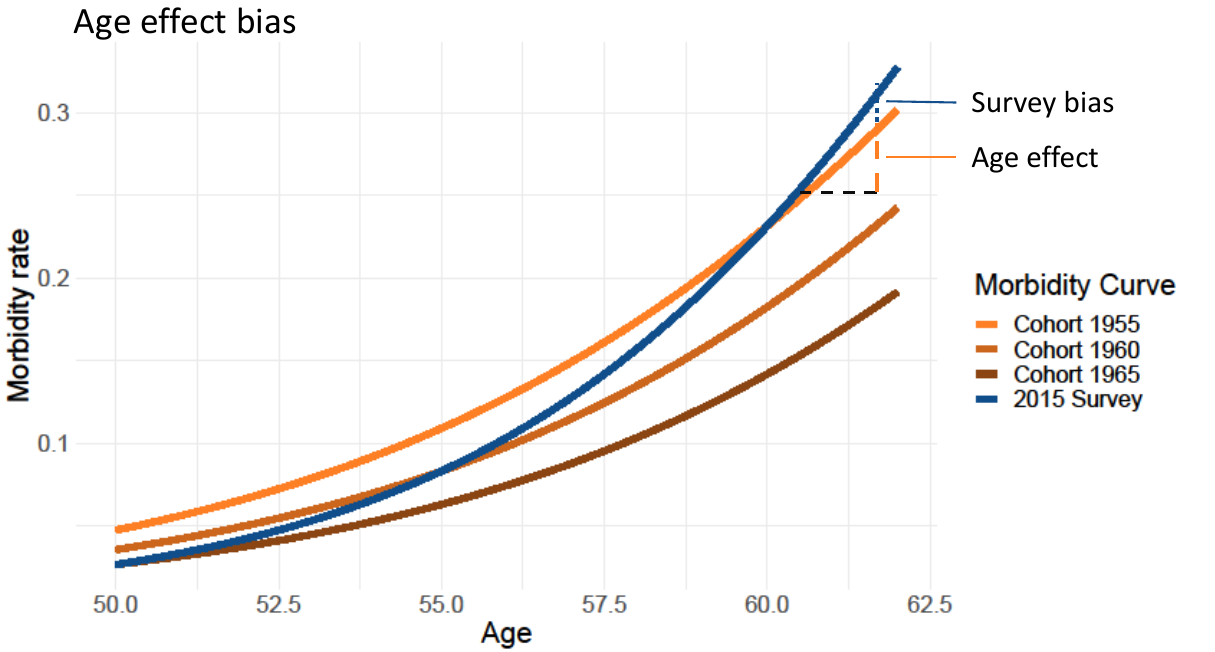}
             \caption{Illustrative example based on synthetic data of the bias one could encounter by defining the morbidity curves under fixed year of survey (blue line), instead of considering a pseudo-panel approach by fixing the birth cohorts (orange lines).}
\end{figure}

Firstly, in contrast to the existing literature \citep[e.g.][]{Pastore-passi22}, we rely on a pseudo-panel approach, considering the birth cohort to which the respondent belongs as a source of temporal dependence, rather than the year of the interview.
{Pseudo-panel approaches are widely used in econometric field to analyse repeated cross-sectional data \citep[see, e.g.][]{deaton1985, moffitt1993, verbeek2008}.} In our application, this approach could reduce the bias in estimating the age effect on disease probability, assuming the presence of non-stationary cohort effects and stationary conditions in the sampling survey scheme. 
Indeed, as illustrated in Figure \ref{fig:bias}, when modelling morbidity curves---i.e. the relationship between age and the probability of a certain disease---using the survey year as time reference could lead to confusing two well-known health risk factors: age and technological (societal and medical) progress. This is because, in a specific survey year, older people are both older and born earlier. 

Secondly, the existing literature often overlooks to account for latent sources of dependence among individuals, which emerge from the dynamic nature of evolving populations. Every person's life journey is shaped by a series of unique historical events, which influence the trajectory of their lives in both time and space. 
As a result, the influence of individual characteristics is likely to fluctuate depending on the contextual circumstances a person encounters or has encountered. In other words, these effects are closely tied to the latent spatial and temporal risk factors that individuals are exposed to.
To address this aspect, we rely on the Bayesian framework, a well-established tool in epidemiology \citep[e.g.][]{DunsonBayesianEp}, to incorporate prior knowledge thoughtfully. In particular, we design the priors of the regression parameters to vary over time and space depending on latent risk factors. 
Modelling local heterogeneity in disease risk through latent spatial risk factor is not new in the field of epidemiology \citep{kelsall2002}. 
However, inspired by the use of additional data on locations  \citep{Best2000JasaPollution, Diggle2010JasaAggregated}, we propose to structure these latent processes over further information recovered by exogenous sources or %aggregating
modifiable risk factors { available at the local health unit level}. % as behavioral patterns and habits, over small homogeneous groups in term of geographical location and temporal context. 
The absence of individual histories in cross-sectional data necessitates the development of suitable methods for inferring causal relationships \citep{wunsch2010, reichenheim2010}. Consequently, BRFSS literature often focuses primarily on assessing the impact of non-modifiable individual risk factors, such as age and sex at birth \citep[see, e.g.][]{AssafJRSSA15, Pastore-passi22}, ignoring valuable information pertaining to modifiable factors.
In the context of health risk analysis, harnessing this information through suitable and sophisticated methods is crucial for obtaining meaningful insights and avoiding potential biases. For instance, let us consider the influence of excessive alcohol consumption on the likelihood of developing chronic diseases. While the medical literature widely recognises the adverse impact of excessive alcohol consumption on chronic diseases \citep{shield2014chronic}, simply incorporating current alcohol consumption as a covariate in a health risk model on cross-sectional data may yield misleading estimates. 
These estimates can inadvertently reflect biases or under-reporting in both health and chronic disease populations \citep[e.g.][]{boniface2014drinking, SubbaramanAlcohol2020}, 
simply because they might reflect the effects of unobserved components (e.g. social inclusion) that positively correlate with a better health condition and, in the extreme case, they might even reflect the presence of an opposite causal relationship in which unhealthy people are forced not to drink.

%As a further element of novelty of our proposal,
Hence, we include modifiable risk factors, such as behavioural patterns and habits, aggregated over small homogeneous groups in terms of geographical location and temporal context to inform the dynamics of the regression coefficients.
%On the other hand, we need a suitable statistical specification to avoid of simply ignore such an information. 

Finally, it is crucial to consider the heterogeneity with respect to different health outcomes, specifically the diverse chronic diseases that contribute to an individual's comorbidity profile \citep{andreella2023}. We propose a multivariate logistic regression specification that takes into account the impact of both individual and environmental risk factors on the marginal probability of developing specific chronic diseases. Additionally, we include a latent comorbidity indicator to capture how these diseases may be further correlated due to an individual’s propensity for comorbidity.

The paper is organised as follows. Section \ref{sec2:data} introduces the behavioural and risk factor surveillance system data that motivate this study. In Section \ref{sec:model}, we present
the statistical methodology, while this will be applied to Italian data
in Section \ref{sec:application}. Finally, Section \ref{sec:discussion} presents some concluding remarks, discussing critically the approach here proposed.

\section{{ Local} and temporal heterogeneity in the BRFSS Italian PASSI data}
\label{sec2:data}

This work is motivated by the richness of information included in the data collected by PASSI \citep[Progressi delle Aziende Sanitarie per la Salute in Italia, translated as ``Advancements of Local Health Units in Italy'',][]{passi}. 
PASSI  is a surveillance system that, since 2008, collects sample data on the behavioural and risk factors of the Italian population in the age span of 18 to 69 years \citep{POSSENTI2021104443, baldissera2011peer}.
Based on the US Behavioural Risk Factor Surveillance, PASSI surveillance system aims to establish a continuously updated and local level database to monitor trends in health issues, risk factors and preventive measures in Italy.
Since public health initiatives in Italy are predominantly organised and assessed at the local level, PASSI engages both regions and local health units directly in the surveillance process, that, supervised by national coordinating group of experts, produces monthly independent cross-sectional samples of the target population.
Annually, information related to approximately 30,000 respondents is collected.
%to produce reports and indicators that describe, at various levels of aggregation, the health and behavioural risks status of the nation (see ISS website, \url{https://www.epicentro.iss.it/passi/}, for the list of produced indicators and their description).
On behalf of local health units, data are gathered by trained workers through a telephone survey. These surveys ask a set of predefined questions of various natures to the respondents, covering their biographical, socio-economic, and health statuses, as well as present and past aspects of their habits and lifestyles.
Personal data, such as sex, age, date of the interview, and local health unit of belonging are collected by default. Aspects regarding their socio-economic status, such as family status and level of education, are inquired about during the interview.
Information on the subjects' health status pertains to their overall well-being and specifically whether certain chronic non-communicable diseases have ever been diagnosed on the day of the interview \citep[see the web page][for details on specific questions]{passi}.
%, while other collected information, such as the perceived status of well-being, will be not considered. 
%Figure \ref{fig:morbidity} reports smoothed morbidity curves
%
 \begin{figure}[ht]
 \label{fig:morbidity}
 \centering
             \includegraphics[width = 0.99\textwidth]{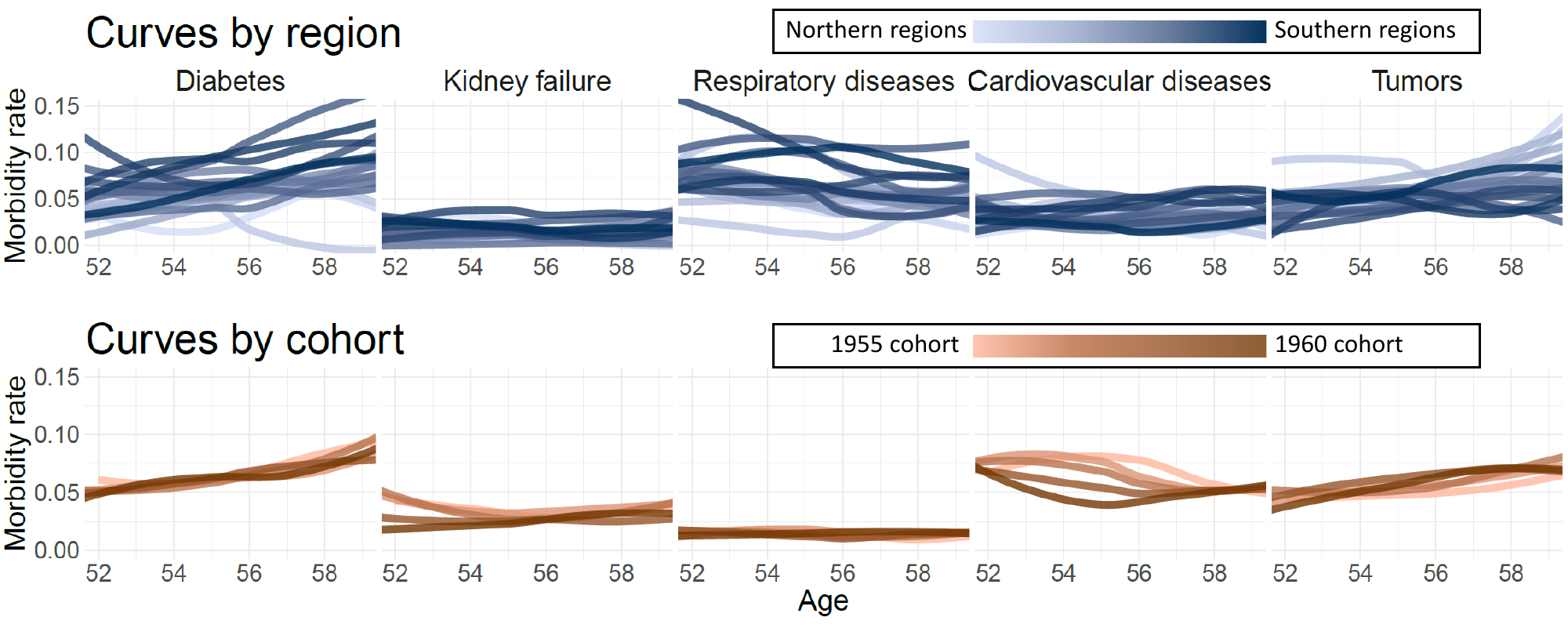}
             \caption{Smoothed morbidity curves representing the relation between age and the frequency of diagnosis for the five considered diseases in the PASSI subjects born between 1956 and 1960 and interviewed between 2008 and 2020.
             Plots in the first row report the curves obtained by grouping by Italian administrative regions (where lighter lines refer to northern regions), while in the second row plots curves are obtained by grouping by cohort (where lighter lines refer to earlier cohorts).}
\end{figure}   
Information regarding the habits and lifestyles of the respondents covers several health-related aspects, such as current dietary habits and alcohol consumption over the past thirty days, smoking behaviors in current and past life, or physical activities pursued during the last $30$ days, among others.
%Other than these, information  concerning habits of taking health screenings and tests are also collected but are not of interest to the present work.

{ 
The focus of this work is on modelling whether some relevant chronic diseases were diagnosed in respondents based on possible risk factors.
Specifically, for a generic respondent $i\;(i=1,\ldots,n_s)$, we target the $n_d$--variate binary response vector $\boldsymbol{y}_i$ indicating the presence or absence of $n_d=5$ diseases, with $y_{ij} = 1$ if the disease $j$ has been diagnosed to the subject $i$.
We take into account the diseases considered by \cite{Pastore-passi22}, namely: diabetes, kidney failure, respiratory diseases (chronic bronchitis, emphysema, respiratory failure), cardiovascular diseases (myocardial infarction, cardiac ischemia or coronary heart disease), and tumors (including leukemias and lymphomas). 
Respondent-specific non-modifiable risk factors, as demographic and socio-economic information, are framed in an $n_p$--variate vector $\boldsymbol{x}_i$ of covariates, with the $h$--th entry $x_{ih}$ indicating the value of covariate $h$ for the respondent $i$.
}

Using the same PASSI dataset, \cite{Pastore-passi22} recently propose a national level Bayesian logistic model in which a binary morbidity indicator, describing whether at least one disease were diagnosed for a single subject, is  regressed against respondent-specific non-modifiable risk factors, namely, age, sex, educational and socioeconomic status. By considering different years of interview, they provide evidence of a morbidity compression in Italy, with respect to risk factors combinations. 
In a similar fashion, \citet{andreella2023} propose a composite indicator that weights disease occurrence differently based on perceived health, with the weights provided by the global burden of disease \citep{salomon2015disability}. Subsequently, they relate the developed index to the non-modifiable risk factors identified by \cite{Pastore-passi22}.

Both the studies focus on a global view of morbidity and on a national perspective. Thus, they do not yet address the local heterogeneity of the specific diseases that is, by nature of Italy, expected. 
Italy is not only a highly variable territory, ranging from coastal areas, to the highest peaks in Europe, but it is also heterogeneous in terms of population composition, socioeconomic status, habits, and exposures to spatial risk factors. 
Thus, if it is reasonable to believe that these aspects influence people's health status, it is also reasonable to assume that differences in risk factors are reflected in heterogeneity in the prevalence of chronic diseases across the country. 
Figure \ref{fig:morbidity} depicts smoothed morbidity curves representing the relationship between age and the frequency of diagnosis, aggregated by regions (above) and birth cohort (below), respectively.
From the figure, not only can we observe differences in sample prevalence among diseases at the aggregated level, but it also illustrates how sample prevalence varies differently in the spatial (regions) and temporal (cohorts) domains.
The choice of spatial aggregation by administrative region (20 in Italy) was made for aesthetic purposes, as representation at lower levels (e.g. local heath units) does not allow for a clear reading of the data, due to the high number of curves to represent (over 100 local health units in Italy). It is worth highlighting that the two representations have different levels of aggregation, so it is recommended to make direct comparisons between the graphs only horizontally, and not between figures above and below.

While it is clear that there is a different degree of heterogeneity in disease prevalence, it is less clear how risk factors influence it, and this motivates our work.
We distinguish here between modifiable and non-modifiable risk factors, as well as between individual and contextual risk factors.
We consider a risk factor to be individual if it regards specific characteristics of the subject.
On the opposite, a risk factor is said to be contextual if it pertains the context in which an individual lives.
Non modifiable risk factors represent individual or contextual characteristics regarding the past, which are no longer modifiable by the person or that are unchangeable in the short term. 
On the opposite, modifiable risk factors represent individual or contextual characteristics that can mutate in the short term.
We consider the variables included in \citet{Pastore-passi22}, i.e. sex at birth, age, educational and economic status, to be individual and non-modifiable risk factors, { framed in the covariate vector $\boldsymbol{x}_i$. An age-sex interaction term is also included in $x_i$, }
together with a dichotomous variable, the smoking status, indicating whether the subject is or has been a smoker in the past.
The inclusion of the smoking status as an individual risk factor deviates from the typical analyses 
%should be acknowledged as a novelty in the
of the PASSI dataset. While the smoking status is in fact modifiable, we categorised it as a non-modifiable risk factor because, given the age range under consideration, it is reasonable to assume that being a smoker reflects information about the individual's history rather than a temporary choice. 
Although there is epidemiological evidence suggesting the elimination, in the long term, of the smoking effect after quitting \citep[see, e.g.][]{toll2014}, we treat former smokers as having an equivalent risk to current smokers. This is due to the absence of precise information on both the time of smoking cessation and the time of diagnosis, preventing us from distinguishing individuals who quit smoking after being diagnosed with a disease or former smokers who quit long before the diagnosis. On the other hand, it is reasonable to think that the effect of smoking is anyway negative on health. Thus, by incorporating possible lower-risk subjects (i.e. former smokers) into the smoking population stratum, we expect conservative estimates---i.e. not greater than the actual magnitude---for the smoking effects on morbidity, 
%which we expect to be smaller than their actual magnitudes.
as we anticipate a positive association.

\begin{figure}[ht]
 \label{fig:maps}
 \centering
             \includegraphics[width = 0.95\textwidth]{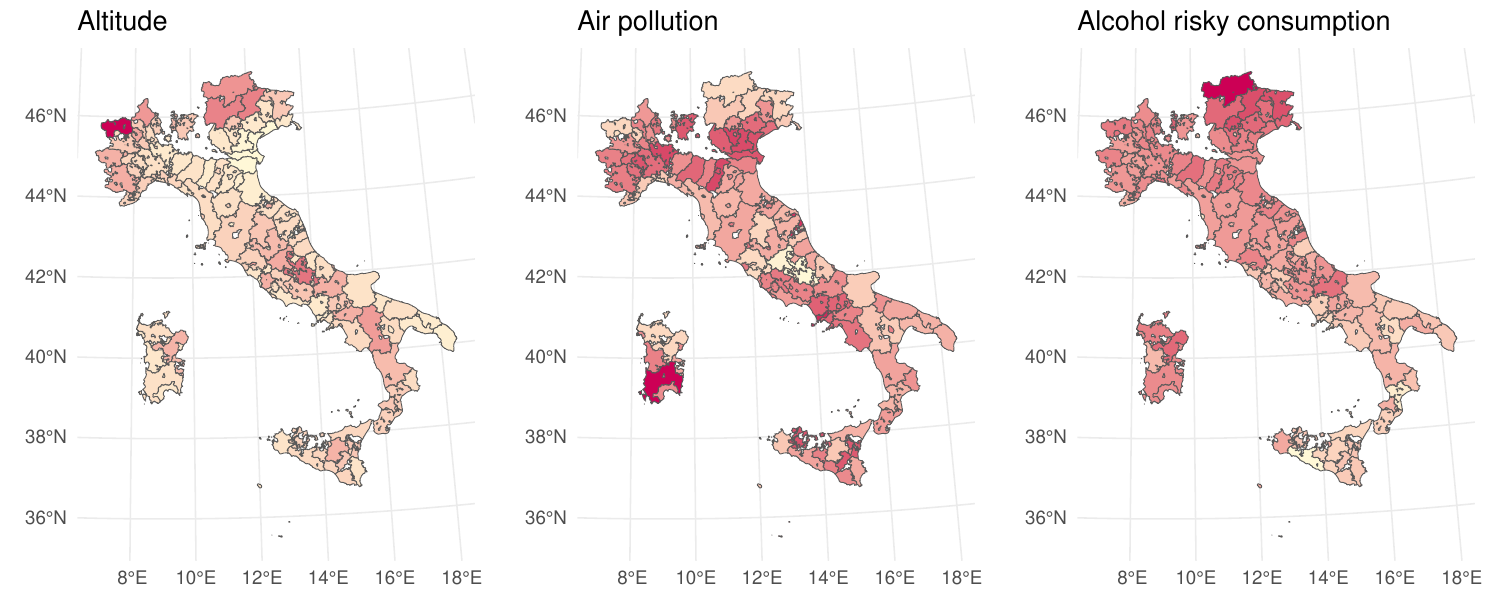}
             \caption{Levels of three possible sources of spatial risk factors in Italy recovered by exogenous data or aggregating modifiable risk factor information available in the PASSI dataset. Darker colours indicate higher levels.}
\end{figure}

Information regarding physical activity, dietary and alcohol consumption habits should be, in principle, considered as individual but modifiable risk factors, because they represent temporary conditions that, potentially, can vary over the course of a lifetime. 
In the literature \citep[see, e.g.][]{smith2007multiple}, it is well known how these factors positively or negatively influence people health status in the long term. 
However, questions regarding these aspects only refer to the last 30 days, and cannot be used to provide a long term and accurate picture of people's past, but rather just a snapshot of their daily habits.
For this reason, we assume that these aspects can provide, if aggregated, a contextual risk factor informing on local and traditional habits and social behaviours.
Although PASSI survey provides rich and useful set of information regarding the habits of the population, it is clear that it can not cover all relevant aspects concerning the context in which one person lives.
For this reason, we decided to address this limitation by considering external data sources that, at the aggregated level, could represent potential contextual risk factors for individual health.
Figure \ref{fig:maps} displays three of the contextual risk factors that we included in our analysis, highlighting the substantial heterogeneity in these aspects across Italy. 
While it is evident why this information could be linked to the health status of the subjects, understanding the specific mechanisms behind this relationship is not straightforward.
{ 

In a typical generalised linear model framework, where the probability of disease $j$ of respondent $i$ is modelled through a linear predictor $\boldsymbol{\beta}_{j}^\top (l_i, c_i) \boldsymbol{x}_i$, we might expect the contextual risk factors of the location $l_i$ and cohort $c_i$ of respondent $i$ to influence the regression coefficient vector $\boldsymbol{\beta}_{j}(l_i, c_i)$.}
How this relation is specified is an important part of the proposed modelling approach, detailed in depth in Section \ref{sec:model} of the paper.
A full description of the additional resources used is reported in the Supplementary Material.

\section{A multivariate logistic regression with { local and temporal} effects}
\label{sec:model}

%For a generic subject $i\;(i=1,\ldots,n_s)$,
%the vector $\boldsymbol{y}_i$ indicates the presence or absence of $n_d$ possible diseases, with $y_{ij} =1$ if the disease $j$ has been diagnosed to the subject $i$. 
%The vector $\boldsymbol{x}_i$ includes the non-modifiable risk factors, with the $h$--th entry $x_{ih}$ indicating the value of covariate $h$ in the subject $i$.

Let Ber$(\pi)$ denote a Bernoulli random variable with mean $\pi$, we specify the logistic model
\begin{equation*}
    y_{ij} \sim \textnormal{Ber}(\pi_{ij}), \quad \pi_{ij} = \textnormal{logit}^{-1}(\eta_{ij}),
\end{equation*}
where $\boldsymbol{\eta}_{i}=[\eta_{i1}, \ldots, \eta_{in_d}]^\top$ is a continuous latent random vector 
\begin{equation*}
    \boldsymbol{\eta}_i = B(l_i, c_i) \boldsymbol{x}_i +\boldsymbol{\gamma}  \epsilon_i, \quad \epsilon_i \sim N(0, \sigma_\epsilon^2),
\end{equation*}
with mean $E(\boldsymbol{\eta}_i) = B(l_i, c_i) \boldsymbol{x}_i$ and residual term $\boldsymbol{\gamma} \epsilon_i$.
In the above specification, $B(l_i, c_i)$ is a $n_d \times n_p$ coefficient matrix with entry $\beta_{jh}(l_i, c_i)$, representing the effect of the $h$--th covariate $x_{ih}$ on disease $j$, as a function of the location $l_i$ and cohort $c_i$ to which the subject $i$ belongs, for $l_i \in \{1, 2, \ldots, n_l\}$ and $c_i \in \{1, 2, \ldots, n_c\}$.
This construction allows one to relate the impact $\beta_{jh}$ of a certain individual risk factor to contextual factors in which the subject lives. 
The scalar $\epsilon_i$ captures part of the residual variability for subject $i$ and represents a measure of the individual propensity to multi-morbidity, which is scaled by the $n_d$--dimensional vector $\boldsymbol{\gamma}$ to regulate disease-specific variability and pairwise associations between them.
Such a construction, after considering the marginal specification of each disease, is equivalent to the latent trait model structure recently proposed by \citet{andreella2023} to define an index of multi-morbidity.
{ To ensure the identifiability of the parameter vector $\boldsymbol{\gamma}$,  we fix $\sigma_{\epsilon}^2=1$ and constrain $\gamma_1$---the first element of $\boldsymbol{\gamma}$---to be positive, while allowing the other entries to vary freely.
}

Since we expect the effects of covariates to vary according to the context in which an individual lives, we assume the entries of ${B}$ to be heterogeneous both in time (cohorts) and space (locations). 
For each scalar coefficient $\beta_{jh}(l_i, c_i)$, we account for temporal dependence by means of a discrete random walk 
\begin{equation}
\label{eq:state-space}
    \beta_{jh}(l_i, c_i) = \beta_{jh}(l_i, c_i-1) + \lambda_{jh}^1 \xi_{jh}^1(l_i, c_i),
\end{equation}
with initial state $\beta_{jh}(l_i, c_0) = \beta_{jh}^0 +   \lambda_{jh}^0 \xi_{jh}^0(l_i),$ where both the shifts $\xi_{jh}^1(l_i, c_i)$ and $ \xi_{jh}^0(l_i)$ are location dependent random variables with standard Gaussian distributed marginals.
Under this representation, $\beta_{jh}^0$ is an unknown scalar parameter for the population mean effect of the earlier cohort $c_0$, and represents the historical mean effect; similarly, $\xi_{jh}^0(l_i)$ represents spatial deviations for location $l_i$, and accounts for the spatial variability observed at the initial cohort $c_0$,
%which is present due to different historical contexts at the location level for cohort , 
with scale $\lambda_{jh}^0>0$ accommodating the scale of the impact of covariate $h$ on the disease $j$; 
finally, the shift $\xi_{jh}^1(l_i, c_i)$, scaled by the value $\lambda_{jh}^1>0$, represents changes in the mean level effect for location $l_i$ between consecutive cohorts, thus representing the latent drivers that lead to different effects in different cohorts. The elements $\lambda_{jh}^0$ and $\lambda_{jh}^1$ are stored in the $n_d \times n_p$ scale matrices $\Lambda^{0}$ and $\Lambda^{1}$, respectively.

To account for dependence between local health units, we assume that both  $\xi_{jh}^0(l_i)$ and $\xi_{jh}^1(l_i, c_i)$ are realizations of zero-mean { Gaussian random variables, defined over set of available locations}, with correlation functions $\mathcal{C}_{jh}^0$ and $\mathcal{C}_{jh}^1$, respectively, depending on contextual risks factors.

We express $\mathcal{C}_{jh}^s$ ($s \in \{0,1\}$) as a convex combination of different functions, by considering the covariance between locations $l$ and $l^\prime$ to be
\begin{equation}
\label{eq:cov}
    \mathcal{C}_{jh}^s(l, l^\prime) = \sum_{m=1}^{n_f}  \omega_{jhm}^s \mathcal{K}_m(l, l^\prime),
\end{equation}
where $\omega_{jhm}^s \in (0,1)$ are weights such that $\sum_{m=1}^{n_f}  \omega_{jhm}^s=1$, and $\mathcal{K}_m(l, l^\prime)$ are functions depending on external information on the locations and  such that $\mathcal{K}_m(l, l)=1$ and $\mathcal{K}_m(l, l^\prime)=\mathcal{K}_m(l^\prime, l) \geq 0$.
As a general requirement, we ask  $\mathcal{C}_{jh}^s$ to be valid covariance functions, for any $j,h,s$ and locations, without necessarily requiring it to $\mathcal{K}_m$.
Although several alternative specifications for the function $\mathcal{K}_m$ might be considered, in this paper, we employ two types of functions. The first one takes the general form of 
\begin{equation}
\label{eq:kernel-reg}
    \mathcal{K}_m(l, l^\prime) = \sum_{r=1}^R \mathbb{I}\{\mathcal{R}_r(l) \} \,\big[ \theta_r \,\mathbb{I}\{ \mathcal{R}_r(l^\prime)\} +(1-\theta_r)\,\mathbb{I}(l=l^\prime)
    \big], \quad \theta_r \in (0,1),
\end{equation}
with $\mathbb{I}\{\mathcal{R}_r(l)\}$ equal to one when the condition $\mathcal{R}_r(l)$ is true, and where the $R$ conditions define a partition of the location set.
Such a construction guarantees $\mathcal{K}_m(l, l)=1$ and allows one to consider {local} dependence that cannot be described { through values on a continuous scale, such
%trough a continuous spatial process on $\mathbb{R}^2$,
as administrative conditions on the locations}. For instance, in the PASSI data discussed above, one may consider the partition induced by administrative Italian regions, such that $\mathcal{K}_m(l, l^\prime) = \theta_r$
for any couple of locations $l \neq l^\prime$ belonging to the same region $r$. 
The parameter $\theta_r$ represents the proportion of variance explained by the condition $\mathcal{R}_r$ with respect to the location idiosyncratic variance.
{ 
Alternatively, it is possible to include simple neighborhood structures in the model such as the spatial moving average model \citep[see, e.g.][]{SpatialMovingAverage}, in which 
$\mathcal{K}_m(l, l^\prime) = \theta_{ll^\prime}$ if the locations $l$ and $l^\prime$ share their boundaries, and $0$ otherwise (when $l \neq l^\prime$).
}

The second type of kernel takes the form of
\begin{equation}
\label{eq:kernel-dist}
    \mathcal{K}_m(l, l^\prime) =  \exp\{ -%\alpha_m 
    \mathcal{D}_m(l, l^\prime)\}, 
    %\quad \alpha_m >0,
\end{equation}
where $\mathcal{D}_m(l, l^\prime)$ is a suitable distance function underlying the definition of a corresponding spatial contextual risk factor, {computed considering available information on the locations}.
{ 
In the application section, we consider distances using different data sources, considering information on air pollution, habits, and other geographical information at the local health unit level. 
However, this formulation can also accommodate typical kernel structures used in spatial data analysis.
For instance, plugging in the Euclidean distance between location centroids, in term of their latitude and longitude, would allow for the description of spatial dependence using an exponential covariance function.
While various kernel functions could be useful in practice, providing an exhaustive specification of all possibilities is beyond the scope of this work. We refer to the Supplementary Material for some relevant examples and to specialized literature on multilevel hierarchical models for spatial analysis \citep[e.g.][]{banerjee2003hierarchical} for a detailed discussion.
}

{
 It is relevant to highlight that, } under this general formulation, it is easy to verify that both $\xi_{jh}^0(l_i)$ and $\xi_{jh}^1(l_i, c_i)$ can be thought as linear combinations of two types of independent {multivariate Gaussian random variables}:
the first one correspond to a zero-mean { random variables} with independent steps over sets of locations that fulfill conditions $\mathcal{R}$; the second one correspond to a zero-mean { random variables with correlation that is a} decreasing function of the distance $\mathcal{D}_m$,  describing a vanishing {correlation} between locations that differ in terms of contextual risk factors.
As a consequence, $\xi_{jh}^0(l_i)$ and $\xi_{jh}^1(l_i, c_i)$ can be interpreted as the results of a linear combination of $n_f$ independent risk factors, { varying in the spatial domain of locations}, weighted by means of $\sqrt{\omega_{jhm}^s}$, so that $\omega_{jhm}^s$ represents the importance of the $m$--th factor in determining the marginal variability of $\xi_{jh}^s$. 
%See the Supplementary Material for the detailed derivation of the processes $\xi_{jh}^{0}(l_i)$ and $\xi_{jh}^{1}(l_i, c_i)$ as a sum of latent spatial processes.

As a general comment, our assumption is that similar locations in terms of contextual risk factors are characterised by strong positive association in the deviation or changes of the individual covariate impacts.
{

From a practical perspective, incorporating information regarding contextual risk factors helps capture the correlation expected among similar locations. In this setting, it is of paramount importance including risk factors that effectively reflect the existing variability between locations, which typically exhibit strong spatial dependence.
By doing so,  we propose  a multilevel hierarchical model \citep{gelman2007data} with priors that are not exchangeable for groups of individuals living in different local health units.
%In such a case, well selected risk factors should reflect, at least approximately, variability that can be captured by alternative spatial and random coefficients models, not informed by the available exogenous information. However, these alternative have the drawback of  not leveraging the benefit of borrowing strength provided by the exogenous available information. 
If the included information captures the variability of spatial effects, additional spatial structures to capture local dependence may not be necessary. 
Conversely, capturing the residual spatial variability could be beneficial. 
Our approach differs from standard spatial models for grouped data, which focus on modelling prior precision matrices \citep[e.g.][]{banerjee2003hierarchical}.
By additively defining correlation matrices, 
our proposal offers advantages in terms of interpretation and prior elicitation, as it directly addresses pairwise dependence between locations rather than conditional independence.
}

\section{Application to the chronic diseases in the Italian population}
\label{sec:application}

%\subsection{Data configuration { and alternative models}}
\subsection{ { Model configuration}}

\label{sec:data-config}
We applied the model described in Section \ref{sec:model} to the data collected by the PASSI system between 2008 and 2020. 
We focus our study on subjects interviewed when they were in the sensitive age range 51--62 years old, i.e. in which a chronic condition typically starts \citep{Pastore-passi22}. 
In accordance with this, we take into account only subjects belonging to the 1956--1960 birth cohorts to have sufficient observations in the entire age span. Coherently with the works by \cite{Pastore-passi22} and \cite{andreella2023}, data after the COVID outbreak are excluded.
While the study of this phenomenon would be certainly interesting from an epidemiological point of view, it is outside the scope of this paper and might require some adjustments to both the comorbidity component and the latent process of the model, as well as to accurately consider changes in the survey's questions.

{ As mentioned above, we consider $n_d=5$ diseases---diabetes, kidney failure, respiratory diseases, cardiovascular diseases, and tumors---and six non modifiable risk factors as covariates---age, sex at birth, age-sex interaction, educational, economic and smoking status.}
The addition of the intercept term leads to $n_p = 7$.
Age is scaled according to $(\texttt{age}-51)/11$, where $51$ is the minimum age considered and $11$ is the age span.  
In view of the short age range considered, we assume a linear approximation of the age effect on the mean response in logit scale.

In regard to the latent spatial factors, we define $n_{f} = 5$ kernel structures.
The first kernel $\mathcal{K}_{1}(l, l^\prime)$ is structured as in equation \eqref{eq:kernel-reg}, where the conditions $\mathcal{R}_r(l)$, with $r=1,\ldots,21$, indicate whether location $l$ is included or not in the $r$-th administrative region 
\begin{equation*}
    \mathcal{K}_{1}(l, l^\prime) = \begin{cases}
        1 \quad \text{if} \; l = l^\prime,\\
        \theta_r \quad \text{if} \; l\neq l^\prime \; \text{and $l,l^\prime$ belong to region $r$,}\\
        0 \quad \text{otherwise}.
    \end{cases}
\end{equation*}
As a consequence, the parameter $\theta_r$ is a measure of the level of association between locations in region $r$, while the product $(1-\theta_r)\,\omega_{jh 1} $ represents the proportion of variance explained by a location idiosyncratic error term. 
%for the locations in region $r$.
{ 
The second kernel incorporates information about the neighbouring structure between LHUs by considering a dependence structure induced by a first-order stationary Spatial Moving Average (SMA) process on irregular lattices \citep[see, e.g.,][and the Supplementary Material]{SpatialMovingAverage, SpatialMAsecond}.
This kernel reduces to
\begin{align*}
    \mathcal{K}_2(l, l^\prime) = \begin{cases}
        1 \quad \text{if } l = l^\prime,\\
        \theta_c \,(n_l \, n_{l^\prime})^{-1/2} \quad \text{if } l  \text{ and } l^\prime  \text{ are neighbors},\\
        0  \quad \text{otherwise,}
    \end{cases}
\end{align*}
for $\theta_c \in (0,1)$ and with $n_l$ representing the number of neighbours of location $l$. Under this construction, $\theta_c$ is a contiguity parameter representing the degree of residual dependency shared among neighbouring LHUs. 
}

The other three kernels take the form defined in equation \eqref{eq:kernel-dist}, 
and are based on three corresponding distance functions $\mathcal{D}_m(l,l^\prime)$ underlying continuous contextual risk factors described below.

{
Geographical risk is modelled by one factor based on the Euclidean spatial distance of the scaled average altitude and population density of each Local Health Unit.

The absolute differences between the average PM10 levels across locations in 2021 are used to define the distance underlying an air pollution risk factor.

Finally, a synthetic distance measure is calculated based on information collected by the PASSI survey related to possible individual modifiable risk factors and aggregated by location. 
Specifically, for each Local Health Unit, we consider the proportion of respondents exhibiting risky alcohol consumption, the proportion of respondents that report BMI (Body Mass Index) measures which define them as overweight, and the proportion of respondents with regular consumption of vegetables and fruits. Additionally, we compute the average level of engagement in sporting activities for each Local Health Unit, where each respondent is classified as having low ($0$), medium ($1$), or high ($2$) habits. The Euclidean distance based on these four variables provides the location distance matrix used to define spatial risk in terms of habits.

As explained in Section \ref{sec:model}, in this construction, contextual risk factors represent information that may be related to the observed differences in disease occurrence in Italy. However, it is likely that the included information does not fully capture the existing variability, so the kernel $\mathcal{K}_2$ allows us to account for any residual dependence present among different LHUs. Further details on the definition of the three distance matrices, as well as the data sources, are provided in Section S1 of the Supplementary Material. It is also worth considering the exploration of alternative distances to induce different latent spatial risk factors.
}

\subsection{Prior elicitation}
Given the data configuration, we carefully set the prior distributions on the unknown parameters of the model introduced in Section \ref{sec:model}. %Namely,  $\beta_{jh}^0$ stored in the $n_d \times n_p$ matrix $B^0$, $\boldsymbol{\gamma}$, $\lambda_{jh}^0$, $\lambda_{jh}^1$ stored in the $n_d \times n_p$ matrices $\Lambda^0$ and $\Lambda^1$, respectively, and the terms $\boldsymbol{\omega}_{jh}$ ($j=1,\ldots,d$ and $h = 1,\ldots,p$).
%and $\alpha_m$ ($m=1,\ldots,n_f$).

To model comorbidity, 
%For identifiability purposes, 
we set on $\gamma_1$---the first element of $\boldsymbol{\gamma}$---a half normal distribution prior to constrain it to be positive.
The other elements of $\boldsymbol{\gamma}$ are  distributed a priori according to a multivariate standard Gaussian $N_4(0,I_4)$.

Our application data are characterised by only five observed cohort time points $c_0,\ldots, c_4$, motivating a simplification of the state space model in equation \eqref{eq:state-space}, by setting constant the shift for any cohort  $\xi_{jh}^1(l_i,c_i) = \xi_{jh}^1(l_i)$.
Hence, trajectories of the state space coefficients $\beta_{jh}$ evolve linearly over time according to the model
\begin{equation}
\label{eq:linear}
\beta_{jh}(l_i, c_i) = \beta_{jh}^0 + \lambda_{jh}^0 \xi_{jh}^0(l_i) + \lambda_{jh}^1 \xi_{jh}^1(l_i) (c_i - c_0),
\end{equation}
where $c_0 = 1956$ is the first cohort considered. 
To avoid overfitting, we propose a sparse representation of the  $n_d \times n_p$ matrix $B^{0}$ that stores the elements $\beta_{jh}^0$.
{ Following recent literature in Bayesian factor models \citep{bhattacharya2011, bhattacharya2015}, we specify an over-parameterised factorisation combined with a shrinkage prior on the factorising matrices to control the true rank $r^*$ of the signal in $B^0$.}
Specifically, we set $B^0 = \Phi_B \Delta_B \Psi_B$,
where the generic element $\delta_{rr}$ of the $\text{min}(n_d, n_p)$-dimensional diagonal matrices $\Delta_B$ is a gamma-distributed random variable $\text{Ga}(a_\delta, b_\delta)$ with $E(\delta_{rr}) = a_\delta/b_\delta$ and $\text{var}(\delta_{rr})=a_\delta/b_\delta^2$. 
{ The introduction of the diagonal matrix is equivalent to the core tensor of the so called CP parameterisation \citep{kolda2009} in tensor factorisation model and is also used in matrix rank reduction methods \citep[see, e.g. ][]{schiavon2024}.}
Gamma hyper-parameters $a_\delta = 0.3$ and $b_\delta = 0.6$ are chosen to favour shrinkage, { such that the true rank $r^*$ is controlled by the number of non shrunk elements on the diagonal of $\Delta_B$.}
Prior elicitation on $B^0$ is completed by defining independent standard Gaussian priors on the elements $\phi_{jr}$ and $\psi_{rh}$ of $\Phi_B$ and $\Psi_B$, respectively. 
{
Considering the single entry 
$\beta_{jh}^0$, the model induced is the sum
$\sum_{r = 1}^{\min\{n_d, n_s\}} \psi_{rh}^*$, with $\phi_{jr}^* = \phi_{jr} \delta_r^{0.5}$ and $\psi_{rh}^* = \psi_{rh} \delta_r^{0.5}$ having normal-gamma marginal priors \citep{brown2010}. 
}

Suitable shrinkage priors for avoiding overfitting were also set for coefficients describing the local deviations from national effects. 
{
Specifically, we used a gamma prior $\textnormal{Ga}(\alpha_\lambda^s, b_\lambda^s)$ on each element $(\lambda_{jh}^s)^2$  to induce a normal-gamma prior \citep{brown2010} on each marginal local deviation defined by $\lambda_{jh}^s \xi_{jh}^s(l_i)$.
To determine the hyperparameters, we follow \citet{brown2010} aiming for a balance between flexibility and shrinkage behaviour in various applications of the model. 
We specify an exponential distribution with mean $1$ for $\alpha_\lambda^s$ and set the rate $b_\lambda^s$ equal to  $\alpha^s_\lambda/(\sigma^2_\zeta \rho^s)$, where $\sigma^2_\zeta$ is the variance of the local deviation coefficients $\zeta_{jh}^s$ and $\rho^s$ is the expected proportion of non-shrunk entries of $\Lambda^s$.
We set $\sigma^2_\zeta=0.5$ and $\rho^s = 1/n_p$.}
\begin{comment}
    we specify a spike and slab prior distribution on the elements of $\Lambda^s$, for $s \in\{0,1\}$. 
In particular, we set 
\begin{equation*}
    \lambda_{jh}^{s}\sim \rho \text{Ga}(2, 4) + (1-\rho) \text{Ga}(0.2, 4),
\quad \rho \sim \text{Be}(3,6),
\end{equation*}
with the spike distributed as a gamma $\text{Ga}(0.2, 4)$ to favour the MCMC mixing.
\end{comment}
In this prior settings, for disease-covariate combinations where data do not suggest presence of spatio-temporal variability, cohort-location specific coefficients are shrunk toward $B^0$. On the other hand, this structure allows the model to learn parameters of $\mathcal{C}_{jh}^s$ from the disease-covariate combinations that depend on the contextual risk factors.
For instance, { to learn the function $\mathcal{K}_1(l,l^\prime)$ and $\mathcal{K}_2(l,l^\prime)$ from the data, 
we impose a sufficiently flat beta prior Be$(2, 2)$ centred on $0.5$ on the regional association weights $\theta_r$ and on the contiguity parameter $\theta_c$.}

We set the weights $\boldsymbol{\omega}_{j}= \boldsymbol{\omega}_{jh}$ constant for each regression coefficient $h=1,\ldots, n_p$.
To accommodate the sum constraint, we specify a Dirichlet prior  $\boldsymbol{\omega}_{j} \sim \textnormal{Dir}(\boldsymbol{a_\omega})$,
%\citep[see, for instance, the arguments in][about the use of Dirichlet prior in sparse vector elicitation]{bhattacharya2015}
with $\boldsymbol{a_\omega} = (2,2,2,2,2)^\top$.
{ This prior specification induces a full correlation matrix between locations, as represented in Figure S1 in the Supplementary Material.}
%small enough to guarantee sufficient mass of prior probability on low-entropy $\boldsymbol{\omega}_{jh}$ configurations. %For the sake of simplicity, one may assume that the covariance structure in equation \eqref{eq:cov} is constant for every coefficient $h$, such that $\boldsymbol{\omega_{jh}} = \boldsymbol{\omega_{j}}$ for any $h = 1,\ldots, n_d$.

{ To better clarify the prior induced on the joint distribution of the linear predictor $\eta$ we remind to Figure S4 of the Supplementary Material.
}

\subsection{ Model validation}
\label{subsec:mod-ass}

To approximate the posterior distribution, we run a Hamiltonian MCMC for $3000$ iterations, discarding the first $1500$ iterations. 
%Then, we thin the Markov chain, saving every $5$-th sample. 
The code was implemented in {\sf Stan} and {\sf R} and is available in the Supplementary Material.
{ 
Model assumptions were validated using predictive checks tools for discrete data \citep{predictiveChecksDiscrete}. 
Specifically, observed disease prevalences, computed for each LHU, were compared to those simulated from the posterior predictive distributions. Predictive check plots and Bayesian p-values \citep[e.g. ][]{PValuesMeng, GelmanPvalues} are provided in Section S4 of the Supplementary Material.
The plots show predictive prevalence prevalence centred around observed prevalence for most LHUs, 
with only a few small Bayesian p-values, indicating no substantial deficiencies in the model.
}

\begin{table}
    \caption{ \label{tab:loo} 
     {
     Comparison of predictive capacities of different models using LOO-IC and WAIC 
     differences relative to the best model.
     Standard errors of the differences are reported in parenthesis.
     The models are ranked in descending order based on LOO-IC, with the last column reporting their computational burden. 
    %   Computational burden (last column) is estimated using the Stan implementation of the models, on a Windows machine with an AMD EPYC 7413 24-Core Processor with                  2.65 GHz. 
    }}
    \fbox{%
    \footnotesize
    \begin{tabular}{l|cc|c}
    Model &  $\Delta$ LOO-IC (s.e.) &   $\Delta$ WAIC (s.e.)  & Seconds/Iter \\
    \hline
        Full ST  & - & - & $\sim73$\\
        Full NS ($\omega_{jh2}=0$)  & $6.09$ $(4.25)$ & $6.39$ $(4.25)$ & $\sim64$\\
        Full NST ($\omega_{jh2}=0$, $\Lambda^1 = 0$) &$13.17$ $(7.89)$ & $13.99$ $(7.89)$& $\sim57$\\
        IL ($\mathcal{C}_{jh}^s=I_{n_l}$) & $16.34$ $(12.11)$ & $15.97$ $(12.11)$ & $\sim56$\\
        FE ($\Lambda^0 =\Lambda^1 = 0$)  & $117.59$ $(26.80)$ &  $120.22$  $(26.80)$ &  $\sim40$ \\ 
    \end{tabular}}
\end{table}
Although the use of a complex hierarchical model may need some cautions, it presents some relevant advantages in terms of fitting data.
{
To evaluate these potential benefits, we compare the predictive capacity of our proposed model (Full ST) with those of alternative specifications that simplify the model construction in terms of hierarchical structure and temporal dynamics, thereby reducing computational demands.
Specifically, we assess the following alternative models, listed in descending order in Table \ref{tab:loo}.
The first alternative (Full NS) shares the same covariance structure as our proposed model but assumes no residual spatial variability by setting the weight $\omega_{jh2}$ of the SMA kernel to zero. In this case, the exogenous information in the other kernels is expected to capture the spatial effects, making additional spatial structures unnecessary
The second model (Full NST) further simplifies by excluding both the SMA kernel and temporal dynamics, with $\Lambda^1$ set to zero.}
The third alternative (IL) assumes locations are a priori independent. In this model, $\Lambda^0$ and $\Lambda^1$ are free, but the correlation structures $\mathcal{C}_{jh}^s$ are set to identity matrices. Using an independent specification for locations allows the coefficients to vary more freely in relation to the observed outcome. However, this approach does not allow locations to borrow information from one another.
Finally, we consider a fixed effect multivariate model (FE), where spatio-temporal variability is neglected by imposing $\Lambda^0 = \Lambda^1 = \boldsymbol{0}$, resulting in regression coefficients that remain constant regardless of cohort or location differences.

{
The comparison is conducted using state-of-the-art metrics for evaluating {\sf Stan} models, as implemented in the \texttt{loo} {\sf R} package \citep{looR}. Specifically, we examine the differences in LOO-IC (Pareto-smoothed importance sampling Leave-One-Out Information Criterion) \citep{vehtari2017} and WAIC (Widely Applicable Information Criterion) \citep{watanabe2010} between our proposed model (Full ST) and alternative specifications, with positive differences indicating worse predictive capacity. Both criteria aim to approximate the expected log pointwise predictive density for a new dataset, providing a measure of predictive performance that balances goodness-of-fit with model complexity, albeit through different approaches.
LOO-IC approximates leave-one-out cross-validation of the log pointwise predictive density using importance sampling to estimate the predictive density $p(y_i \mid y_{-i})$. We employ the recent methodology proposed by \citet{vehtari2017}, which fits a Pareto distribution to the upper tail of the importance weight distribution, reducing the variance of the weights and yielding a more stable LOO estimate. In contrast, WAIC \citep{watanabe2010} estimates the expected log pointwise predictive density by calculating the metric within sample and penalising it by the estimated effective number of parameters, which reflects the model's complexity. \citet{vehtari2017} demonstrated that these two criteria are asymptotically equivalent to leave-one-out cross-validation, which explains the similar results obtained.
Both metrics indicate an overall improvement} when moving from models that do not account for spatio-temporal dependence to those that do. Additionally, incorporating a hierarchical structure that leverages exogenous knowledge appears to enhance model fit, { as does the inclusion of a spatial moving average process to capture residual variability. However, the standard errors of the differences, reported in parentheses, suggest that the evidence for improved predictive capacity is not particularly strong when comparing the more sophisticated model with those that already account for local-temporal variability.
Therefore, the increased computational demand, shown in the last column of the table, may not always be justified. Computational burden is reported as the time in seconds required for an MCMC iteration using the {\sf Stan} implementation of the models on a Windows machine with an AMD EPYC 7413 24-Core Processor at 2.65 GHz.

Nevertheless, while the improvement in predictions offered by our proposed model over simpler specifications may not be substantial, it provides a more detailed and insightful interpretation of the phenomenon, yielding quantities of interest that will be analysed and discussed in the next section. The model's generality also makes it promising for alternative applications.
}

\subsection{Application results}
\label{sec:app-res}

\begin{figure}[ht]
 \centering
 \includegraphics[width = 0.5\textwidth]{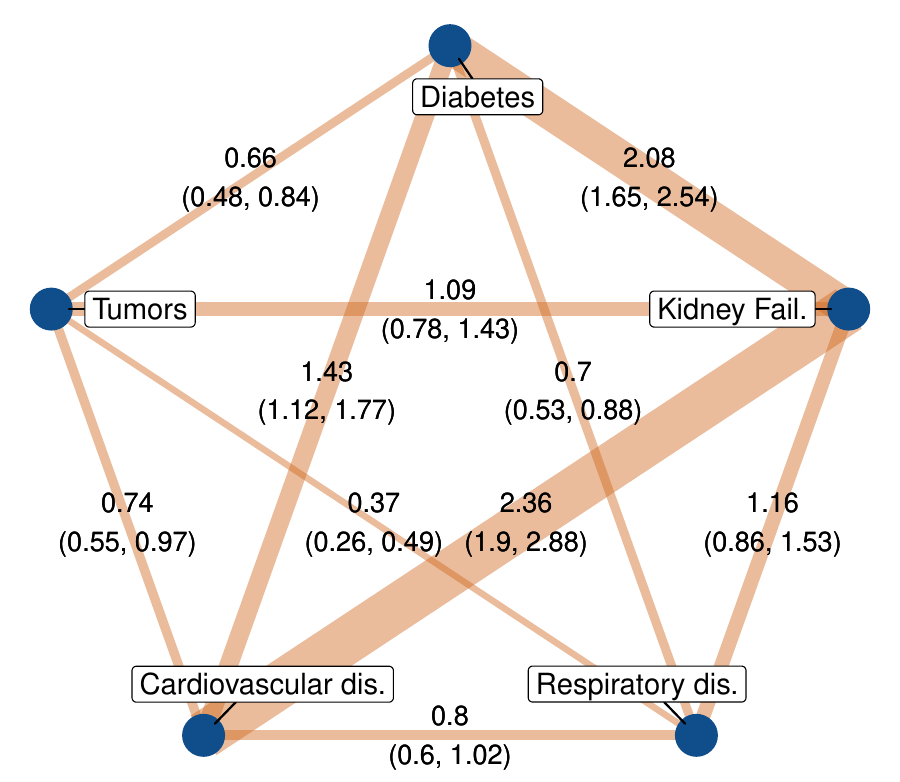}
             \caption{ \label{fig:graph}
 { Adjacency graph between diseases based on a similarity matrix defined as the posterior mean of $\boldsymbol{\gamma}\boldsymbol{\gamma}^\top$. Posterior mean and $95\%$ credible intervals (in parenthesis) are reported on the edges.}
 }
\end{figure}

The inclusion of the co-morbidity vector helps in better capturing residual variability.
Interestingly, all the elements of $\boldsymbol{\gamma}$ are characterised by posterior distribution concentrated on positive values. This means that
%the movement in the direction of multi-morbidity have the same sign. In other words, 
if the individual residual $\epsilon_i$ is positive (or negative), the residual propensity to multi-diagnoses increases (or decreases) for all the diseases considered, with larger changes in absolute value for cardiovascular diseases and kidney failures, as they are characterised by larger value of $\boldsymbol{\gamma}$. Figure \ref{fig:graph} reports a graphical representation of the posterior means of $\boldsymbol{\gamma}\boldsymbol{\gamma}^\top$, in which the five diagnoses considered are connected by edges whose width represents the value $\gamma_j \gamma_{j^\prime}$, with $j \neq j^\prime$.  As a result, this representation allows us to understand how strong is the tendency of the diagnoses to appear (or not appear) together, after controlling for the individual risk factors considered.
{ 
For instance, the strong association observed between cardiovascular disease and kidney failure is well-supported by epidemiological evidence. Chronic kidney disease is a significant risk factor for cardiovascular disease, with a substantial portion of morbidity and mortality in chronic kidney disease patients attributed to cardiovascular complications \citep{foley1998, cai2013}. For a detailed discussion on the epidemiology and pathophysiological mechanisms of the increased cardiovascular risk in patients with chronic kidney disease, refer to \citet{gansevoort2013}.
Similarly is well known and described in the literature the link between diabetes and kidney failure \citep[see, e.g.][]{KOYE2018121} or cardiovascular diseases \citep[see, e.g.][]{kaur2002diabetes}.
}

%On the other side, it is evident the benefit of using a co-morbidity vector to model the individual propensity to diseases. 
%Interestingly, the improvements given by the inclusion of the co-morbidity vector appears to be larger than those given by the inclusion of a suitable spatio-temporal structure, suggesting how the addition of a latent factor in the predictors is crucial to catch individual-specific propensity to disease which is not caught by disease marginal models. This result justifies the use of a multivariate approach with covariance modelling with respect to the simpler marginal models.

\begin{figure}[ht]
 \centering
             \includegraphics[width = 0.95\textwidth]{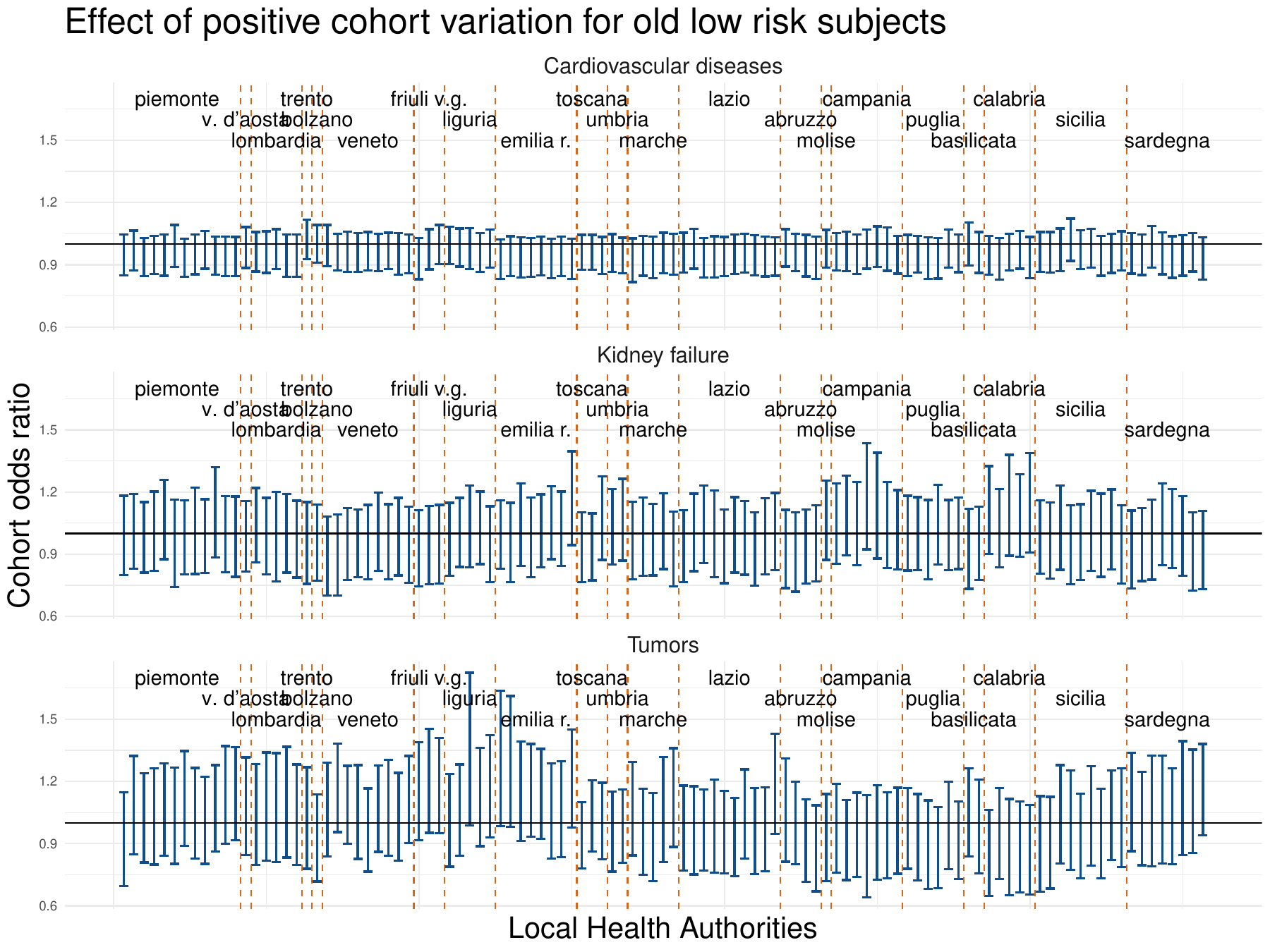}%\\
             \caption{ \label{fig:MC_main}  Posterior $90\%$ credible intervals of the predictive odds-ratio between being born in two subsequent cohorts for a 62 years old low-risk subject in our sample, i.e. a wealthy highly educated female non-smoker. Vertical dashed lines delimit the Italian administrative regions.
             Horizontal line indicates no variations.}
\end{figure}

\begin{table}
    \caption{\label{tab:fe} Mean and $95\%$ credible intervals (in parenthesis) of the posterior distributions of the fixed effect coefficients $B^0$. Bold cells indicate coefficients characterised by credible intervals not including the zero.
    The reference levels for binary covariates are: \textit{female, low educated, low economic status, non-smoker}.
    }
    \fbox{%
    \footnotesize
    \begin{tabular}{lccccc}
 & Diabetes & Kidney fail. & Respiratory dis. & Cardiovascular dis. & Tumors  \\ \hline
(Intercept)  & $ -\mathbf{3.46}$ & $-\mathbf{5.15}$ & $-\mathbf{2.62}$ & $-\mathbf{4.46}$ & $-\mathbf{3.08}$ \\ 
 & $( -3.78, -3.15)$ & $(-5.83, -4.56)$ & $(-2.91, -2.34)$ & $(-4.87, -4.09)$ & $(-3.31, -2.86 )$ \\ 
Sex  & $ \mathbf{0.43}$ & $-0.02$ & $-0.27$ & $\mathbf{0.59}$ & $-\mathbf{1.05} $ \\ 
 & $( 0.14, 0.72)$ & $(-0.58, 0.51)$ & $(-0.55, 0.01)$ & $(0.26, 0.89)$ & $(-1.33, -0.77 )$ \\ 
Edu  & $ -\mathbf{0.49}$ & $-\mathbf{0.73}$ & $-\mathbf{0.30}$ & $-\mathbf{0.32}$ & $0.19 $ \\ 
 & $( -0.68, -0.27)$ & $(-1.12, -0.38)$ & $(-0.45, -0.13)$ & $(-0.62, -0.1)$ & $(-0.06, 0.35 )$ \\ 
Eco  & $ -\mathbf{0.50}$ & $-\mathbf{0.71}$ & $-\mathbf{0.54}$ & $-\mathbf{0.57}$ & $-0.14 $ \\ 
 & $( -0.75, -0.25)$ & $(-1.10, -0.38)$ & $(-0.72, -0.37)$ & $(-0.80, -0.35)$ & $(-0.31, 0.02 )$ \\ 
Smoke  & $ 0.16$ & $0.21$ & $\mathbf{0.61}$ & $\mathbf{0.56}$ & $\mathbf{0.26} $ \\ 
 & $( -0.05, 0.32)$ & $(-0.13, 0.73)$ & $(0.34, 0.8)$ & $(0.34, 0.82)$ & $(0.07, 0.42 )$ \\ 
Age  & $ \mathbf{0.77}$ & $-0.20$ & $-\mathbf{0.53}$ & $0.01$ & $\textbf{0.49} $ \\ 
 & $( 0.16, 1.15)$ & $(-0.83, 0.43)$ & $(-0.87, -0.22)$ & $(-0.42, 0.46)$ & $(0.06, 0.80 )$ \\ 
Age:Sex  & $ 0.31$ & $0.58$ & $0.38$ & $\mathbf{0.67}$ & $\mathbf{0.57} $ \\ 
 & $( -0.07, 0.69)$ & $(-0.02, 1.21)$ & $(-0.04, 0.77)$ & $(0.17, 1.16)$ & $(0.12, 1.01 )$ \\
    \end{tabular}}
\end{table}
In Table \ref{tab:fe}, we report the national effect estimates $B^0$, where the intercepts represent the morbidity risk on the baseline category referring to a $51$ years old non-smoker female, with a low educational and economic status.
The age effect coefficient and its sign varies among diagnoses, an aspect which was not taken into account in other works where a positive age effect has been found when considering a synthetic morbidity index \citep[see, e.g.][]{Pastore-passi22}. 
The reason for this behaviour can be manifold. 
On one hand, it might reflect the inherent heterogeneity of the diseases under consideration, characterised by diverse disease courses and diagnoses. This variability occasionally leads to counter-intuitive behaviours, such as the decreasing age effect observed in respiratory diseases.
While the interpretation of these findings could be of significant interest to clinicians and public health experts, it falls beyond the scope of this study.
%For example, the decreasing age effect on respiratory diseases for women may be related to the high incidence of acute respiratory failure in pregnant women \citep{park2016, lapinsky2017} associated with high mortality rates \citep{vincent2008}.
%or full recover in later years 
On the other side, differences with respect to existing literature reflects both the use of a multivariate morbidity response approach and that our pseudo-panel perspective potentially reduces the bias induced by the presence of non-stationary changes in age-effects between cohorts, as discussed in Section \ref{sec:intro}.
{ As an illustration, the null age effect estimated on the cardiovascular disease probability in female patients, paired with a general slight improvement over female cohorts (refer to Figure \ref{fig:MC_main}), may result in a positive estimate of the age effect if survey year is used as the time reference, as is typically done in cross-sectional analyses \citep{Pastore-passi22, andreella2023}.}
The interpretation of the other parameters is also relevant from both an epidemiological and a policy perspective. 
For instance, the negative effects of Economic Status (Eco. S.) highlight how the well-off population lives in better health conditions, an aspect which is typically reported in the literature \citep{marmot2005, minardi2011}.
Similar observations can be made regarding educational level. However, a higher educational level does not correspond to a reduced number of tumour diagnoses. Given that the response data represent the occurrence of diagnoses rather than the occurrence of disease, the non-negative effect may reflect greater awareness among individuals with higher education about the importance of early diagnosis.

%On the opposite, an higher educational level does not always reflect a reduced number of diagnoses in some diseases.
%Due to the nature of the response data, that represents the occurrence of the diagnosis rather than the occurrence of the disease, the positive effect may reflect an increased awareness of those with higher education with respect to the importance of early diagnosis.% of the disease, such as in case of tumors.

%The full list of parameter estimates is  provided in the Supplementary Material, as well as the parameter estimates of the simplified models... 
%In the latter case, the ... parameters present  lower/greater estimates due to...

\begin{figure}[ht]
 \centering
             \includegraphics[width = 0.95\textwidth]{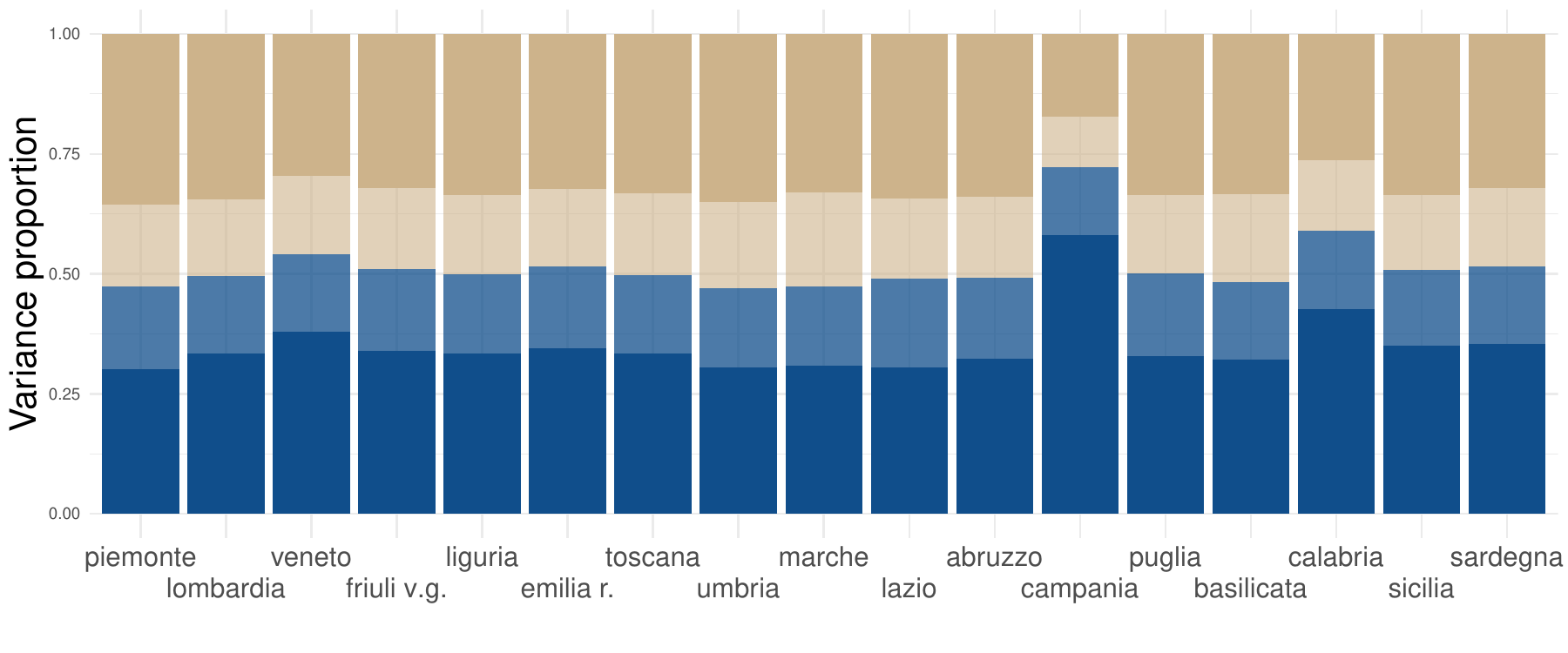}
             \caption{ \label{fig:theta}
 Posterior mean of $\boldsymbol{\theta}$ are displayed by blue bars, while the complementary proportion of local variance is coloured in beige. Posterior interquartile range is shown by lighter bars. Regions including a single local health unit are omitted.}
\end{figure} 
In countries characterised by a degree of local autonomy in the management of health authorities, as is the case in Italy, the ability to focus on local variations aids in identifying characteristic behaviours that are useful for various purposes, including informing policymakers or creating targeted prevention campaigns at regional or local level. 
All contextual risk factors are relevant in explaining local differences, with the administrative regional risk factor playing a crucial role in characterising the local variability of respiratory diseases
%The administrative spatial risk factor is estimated in our model as one of the main driver in explaining local differences, especially for respiratory diseases 
(refer to Figure S2 in the Supplementary Material).
However, the autonomy degrees allowed to local health units within each region may differ.
The parameter $\boldsymbol{\theta}$ displayed in Figure \ref{fig:theta} represents a measure of common variability among locations within each region, with respect to the location idiosyncratic variability. The Campania region stands out as the region with the highest internal consistency. 
%On the other extreme, the presence of a particular city as Rome, very different from the territories of the other local health units,  may be related to the low level of internal consistency in the Lazio region.

\begin{figure}[ht]
 \centering
             \includegraphics[width = 0.99\textwidth]{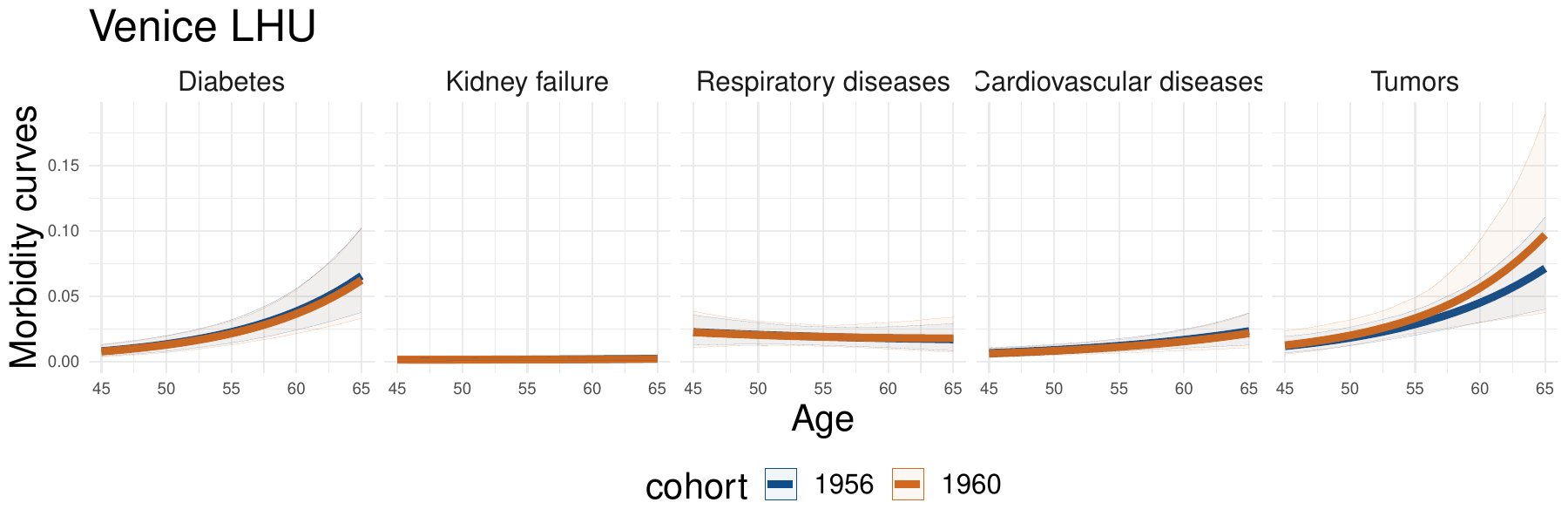}

\caption{            \label{fig:curves}
Example of predicted morbidity curves for five diseases at the local health unit level (LHU of Venice) for wealthy, highly educated and non-smoker male population.}
\end{figure} 
Our model considers a whole set of local factors to help understanding the territorial heterogeneity.
This allows one to study differences in random coefficients among different local authorities or focus on specific one of them.
In this regard, Figure \ref{fig:curves} shows, as example, morbidity curves of Venice health unit for wealthy, highly educated and non-smoker male population. Curves displayed are predicted for two different cohorts, where orange curves correspond to the most recent one. 
We note the two cohorts are characterised by similar behaviours, but for an increase in tumor occurrences. 
This phenomenon may be partially due to improvement in early diagnosis in recent years \citep[see, e.g. the Italian Association for Cancer Research website][]{airc}, as well as ``overdiagnosis'' \citep{welch2010}.
%An interesting aspect can be noticed by looking at the morbidity curves in diabetes: it would seem that there is a slight cohort improvement for wealthy people, suggesting how economic inequalities may be associated to inequalities in health improvements, at least for diabetes in Venice.

\begin{figure}[ht]
 \centering
             \includegraphics[width = 0.95\textwidth]{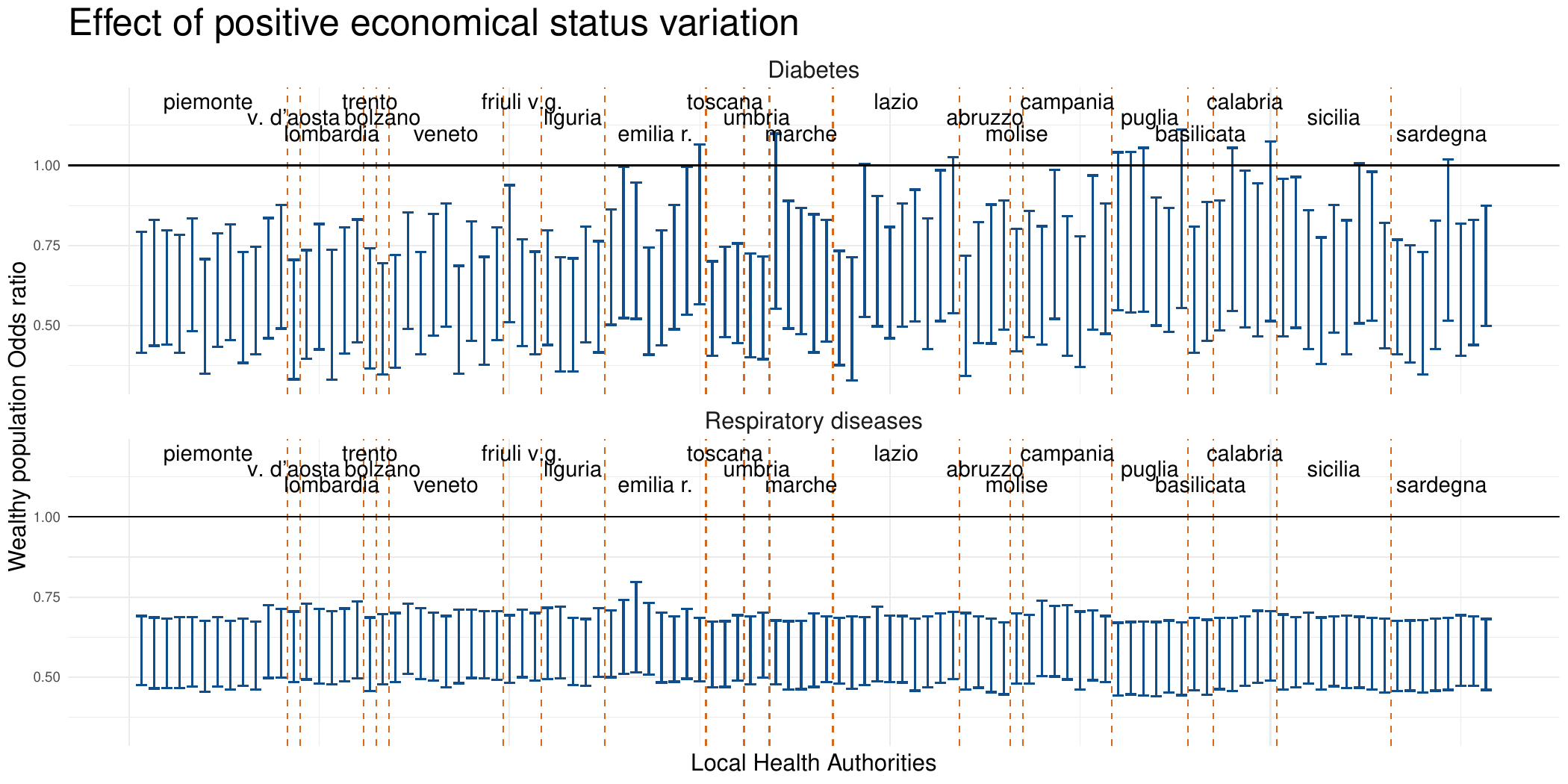}

             \caption{ \label{fig:w_OR} Posterior $90\%$ credible intervals of the economic status coefficient odds-ratio for 1956 cohort and all local authorities considered for diabetes (top) and respiratory diseases (bottom). Vertical dashed lines delimit the Italian administrative regions.}
\end{figure} 
In addition to this, it is interesting to compare the varying effects in different local health units, for each disease and risk factor. 
Figure \ref{fig:w_OR} reports the odds ratios for diabetes and respiratory diseases, derived from the random coefficients of different Italian health units, between people with and without economic difficulties. 
The figure refers to the first cohort considered (i.e. $1956$) and then it should be seen as a representation of local differences accumulated up to $1956$ cohort.
Values below one indicate the reduction factor in the odds of diagnoses when comparing a well-off person with a person with economic difficulties---all other covariates being equal---indicating that, for the diseases considered, being wealthy reduces the probability of diagnoses overall, as previously highlighted. 
Interestingly,  local variability  of the odds ratios (and the relative model coefficients) strongly depends on the disease considered. 
Diabetes coefficients manifest local heterogeneity, both within and between regions, where gaps between wealthy and non-wealthy people in diagnosis seem more evident in northern regions (left part of the panel) than the southern ones (Puglia, Basilicata, Calabria, Sicilia). 
On the contrary, the economic status effects for respiratory diseases do not appear to be affected by local patterns, with the odds ratios that appear to be all concentrated around $0.60$, meaning that economic difficulties affect the diagnosis of this issue approximately in the same way all over Italy. 
Variability measures of differences across local health units are given by the scale matrices $\Lambda^s$ ($s=0,1$), that contain scales to inflates (or deflates) the differences among local health units captured by the correlation matrices $C_{hj}^s$. 
Table S1 in the Supplementary Material reports the posterior means and the credible intervals for all the parameters in these matrices. 
It is notable that entries in $\Lambda^0$ are generally larger than the respective entries in $\Lambda^1$. 
This aspect indicates an higher variability between local health units in terms of historical differences, i.e. differences accumulated up to $1956$ cohort, with respect to the ones that represent the dynamic evolution of coefficients from cohort to cohort, as expected.
In fact, low estimates of $\Lambda^1$ parameters indicate small to negligible temporal variations among coefficients, excluded the age coefficient for tumors, whose posterior mean results to be much larger than the others. 
This result reflects on an higher variability among different locations and on an increased occurrence in tumor diagnoses for younger cohorts, as previously discussed for the Venice case.

In view of the recent literature pointing out the shift toward elder ages of the morbidity curves in Italy \citep{Pastore-passi22, demuru2016}, it is worth asking whether a morbidity compression has being highlighted by our estimates. 
A unique and unambiguous answer cannot be given, as it involves multiple aspects to consider.
First, our flexible specification for multivariate responses highlights how this effect can vary for different combinations of specific segments of local population and diseases.
Second, it is crucial to distinguish the different causes of morbidity variations over years. Improvement in population health conditions may be leaded by general changes in individual risk factors, with several of them that can be influenced by policy makers.
For instance, coefficient estimates of our model suggest that improvements in educational and economic status, as well as the reduction of tobacco consumption, would lead to a general compression of morbidity in younger cohorts due to a different stratification of the population.
On the other hand, our model also provides tools to measure the morbidity compression effect due to technological evolution and reduced environmental risk, by estimating changes in risk factor coefficients across different cohorts. 
{ 
The pseudo-panel perspective adopted in this work allows us to decouple the cohort effect from the age effect resulting in a more straightforward interpretation.
}
Considering the linear dynamic over cohorts defined in equation \eqref{eq:linear}, the cohort variation in a risk factor coefficient can be measured by $\lambda_{jh}^1 \xi_{jh}^1(l_i)$ and remains constant for one-year variations in birth cohort. 
As shown in Figure \ref{fig:MC_main}, the odds morbidity ratio between two subjects born in subsequent cohorts strongly depends on the disease considered. In our example, we considered a 62 years old non-smoker female, with high educational and economic status. 
While the cardiovascular diseases (central panel) seems to be characterised by a generalised technological end environmental morbidity compression over the all the regions of Italy, the bottom panel suggests an increase over cohorts of tumor morbidity in northern regions and Sardinia for the given subjects. More specifically, Liguria and Emilia Romagna present the local health units characterised by the highest increased risk (corresponding to the city areas of Genova, Piacenza, and Parma). 
In the top panel we notice high LHUs' heterogeneity in cohort variation of risk effects for kidney failure disease.
Figure S3 in the Supplementary Material reports the effect of positive cohort variation for the remaining diseases.

\section{Discussion}
\label{sec:discussion}
In response to the recent surge in demand for health monitoring in the population, this paper introduced a novel methodology for analysing repeated cross-sectional data, applied specifically to Italian behavioural and risk factor surveillance data (PASSI).
A multivariate local and temporal logistic model for chronic disease diagnosis was proposed, demonstrating the capability to capture latent local heterogeneity through a dynamic formulation of the regression coefficients. The prior of the covariance structure was informed by exogenous data on local health units and spatially aggregated risk behaviours. Leveraging this additional source of information allowed us to enhance model fitting while retaining the ability to focus on local behaviours, serving various purposes such as informing policymakers or creating targeted prevention campaigns at regional or finer levels.
%This aspect proved crucial for improving out-of-sample predictions, as discussed in Section \ref{sec:app-res}.
The inclusion of a comorbidity vector in the linear predictor enabled the consideration of individual propensities to diseases and the modelling of residual relations among diagnoses. 
The application to the PASSI data illustrated how the advantages of our model specification translated into potentially valuable insights and advice for policymakers and health surveillance systems. 
For instance, we recovered the local morbidity curves and identified both individual and contextual factor risks. 
Additionally, comparing varying coefficients enabled the measurement of technological and environmental risk variations among cohorts by diseases. However, the small range of cohorts considered so far did not permit obtaining definitive results in this sense.

{
 
When using survey data, it is typical the use of sampling weights for more reliable inference \citep[see, e.g.][]{CHEN201433, zhaoSampnlingWeights19}. Sampling weights are computed to reflect each unit's importance in representing population aggregates.
In the considered dataset, most sampling weights are concentrated around one.
Furthermore, they are a function of demographic information, which is already included in the individual predictor $\boldsymbol{\eta}_i$. 
In view of these two aspects, the inclusion of sampling weights provides no significant benefit for our focus.
Nevertheless, the use of sampling weights is crucial when defining population aggregates, such as actual numerical values or disease prevalence at the regional or national level \citep{gao2021improving}. This last aspect is a relevant topic in Bayesian modeling \citep{gelman2007struggles, gelman2024regression}, representing a worthwhile direction for future research.
}

This paper was motivated by the analysis of Italian data spanning from 2008 to 2020 within an age range of 51-62.  
Despite its undeniable advantages, discussed in Section \ref{sec:intro}, the pseudo-panel approach necessitates an extensive surveillance period to analyse long-age curve behaviours effectively.
In our case, constrained by the limited surveillance period in Italy, the analysis concentrated on $22$ thousand respondents, a notably smaller sample compared to those processed by BRFSS systems in other countries (e.g. the U.S.).
Although our model appears promising for analysing such data, the estimation method required adjustments due to its computational intensity and time demands, taking more than a day for $3000$ MCMC iterations. 
In this regard, variational Bayes approaches may represent an efficient alternative worth exploring, and the development of a suitable online monitoring filter could address the continuous updates of the database.
The proposed approach is designed to investigate local behaviours, which necessarily reflect the information gleaned from a small number of respondents, particularly for smaller local health units. 
To address this challenge, we proposed enhancing the estimates by incorporating exogenous information as prior knowledge, such that ``similar'' local health units behave similarly, a priori. This aspect aligns well with the model specification within a Bayesian framework.
Although our method assigned a posterior weight to the external information used, the selection of relevant prior information is crucial. Our proposal demonstrated improvements in terms of out-of-sample predictions, justifying its application.
However, exploring additional exogenous information or alternative methods of incorporating it into the model remains a pertinent aspect for future research, potentially involving collaboration with domain experts.

\section*{Aknwoledgement}
The authors are grateful to the SoSta group  (A. Andreella, M.Bertani, G. Bertarelli, M. Marzulli,  A. Pastore, M. Pittavino, S. Tonellato) at Ca' Foscari University of Venice and to L. Compagno for their insightful comments. We also thank the Gruppo Tecnico PASSI (PASSI national coordination team at ISS) for their valuable feedback, and the PASSI network interviewers for their diligent work in data collection.

\section*{Declarations}

\subsection*{Funding} This manuscript was developed within the project funded by Next Generation
EU - Age-It - Ageing well in an ageing society project (PE0000015), National Recovery and Resilience Plan (NRRP) - PE8 - Mission 4, C2, Intervention
1.3. The views and opinions expressed are only those of the authors and do not necessarily reflect those of the European Union or the European Commission. Neither the European Union nor the European Commission can
be held responsible for them.

\subsection*{Conflicts of interest} The authors declare that they have no conflict of interest.

\subsection*{Availability of data and material} 
 PASSI surveillance data can be accessed at \url{http://www.epicentro.iss.it/passi/}. The dataset used for the analyses is not publicly available due to specific policy of the National Institute of Health and of the Italian Ministry of Health, but it is available by the National Institute of Health upon reasonable request.

\subsection*{Code availability} { The {\sf R} and {\sf Stan} code used in this study is available in the Supplementary Material, along with a synthetic dataset for reproducible examples.}

\bibliographystyle{rss}
\bibliography{references}   % name your BibTeX data base

%%%%%%%%%%%%%%%%%%%%%%%%%%%\begin{thebibliography}{}
%%%%%%%%%%%%%%%%%%%%%%%%%%%\bibitem{RefJ}
%%%%%%%%%%%%%%%%%%%%%%%%%%%%Author, Article title, Journal, Volume, page numbers (year)
%%%%%%%%%%%%%%%%%%%%%%%%%%%%\bibitem{RefB}
%%%%%%%%%%%%%%%%%%%%%%%%%%%%Author, Book title, page numbers. Publisher, place (year)
% etc
%%%%%%%%%%%%%%%%%%%%%%%%%%%%%%\end{thebibliography}

\newpage

\end{document}